# Surface structures of graphene covered Cu (103)

Yui Ogawa[1,*], Yuya Murata[2], Satoru Suzuki[1,†], Hiroki Hibino[1,3], Stefan Heun[2], Yoshitaka Taniyasu[1], and Kazuhide Kumakura[1]

[1] *NTT Basic Research Laboratories, NTT Corporation, 3-1, Morinosato Wakamiya, Atsugi, Kanagawa 243-0198, Japan*

[2] *NEST, Istituto Nanoscienze-CNR and Scuola Normale Superiore, Piazza San Silvestro 12, 56127 Pisa, Italy*

[3] *Kwansei Gakuin Univ., 2-1, Gakuen, Sanda, Hyogo 669-1337, Japan*

*† Present address: Laboratory of Advanced Science and Technology for Industry, University of Hyogo, Kamigori Hyogo 678-1205, Japan*

E-mail: yui.ogawa.er@hco.ntt.co.jp

We studied the surface structures of chemical vapor deposited (CVD) graphene on Cu(103). The graphene covered Cu surface had (103) facets parallel to the Cu [010] direction, on which triangular patterns were formed. In contrast, the bare Cu surface showed no facets. Post-growth thermal annealing in an ultra-high vacuum induced surface changes on the Cu (103) facets. The reorganization of the Cu surface by the post–growth thermal annealing led to a change in the lattice strain and hole doping level of the CVD-grown graphene.

Chemical vapor deposition (CVD) growth on copper (Cu) surfaces, especially on commercial Cu foil, is a promising method for obtaining large-area graphene.[1, 2] Cu acts not only as a substrate for graphene growth but also as a catalyst to decompose hydrocarbon gas sources such as methane ($CH_4$). The low solubility of carbon in Cu is helpful for single-layer graphene growth.[1, 3] The crystalline nature of Cu is also relevant to the properties of graphene. For instance, the nucleation and crystal orientation of CVD-grown graphene depends on the crystal orientation of the surface.[4-8] The orientation of the Cu surface also affects charge doping and lattice strain in graphene because the work function and crystal relationship, which are associated with symmetry matching and superlattices, vary with it.[9-14] The previous reports studied mainly CVD growth of graphene on low Miller index planes such as Cu (111), (001), and (101), not higher index planes.

The CVD-grown graphene typically has structural undulations, including ripples and wrinkles, as well as lattice strain due to the different thermal expansion coefficients of graphene and Cu.[15] The undulations degrade physical properties, such as electrical carrier mobility, mechanical strength, and thermal conductivity after the graphene has been transferred onto the dielectric substrates.[16-18] Post-growth annealing has been proposed as a way to smooth out the structural undulations.[19,20] It is therefore important to understand the effects of thermal annealing on the surface morphology of graphene grown on Cu, because the surface may reorganize during annealing to reduce the interface energy between the Cu and graphene.[9, 14, 19] Therefore, gaining an understanding of how annealing can be used to control surface structures of graphene after growth is an important challenge for engineering of the graphene properties.

Here, we studied CVD-grown and post-growth thermally annealed graphene on Cu foil with a high index (103) plane. The surface structure, lattice strain, and carrier doping of as-grown and thermally annealed samples were characterized.

Samples were prepared in three steps: pre-growth thermal annealing of the Cu foil, CVD growth of graphene on the Cu foil, and post-growth thermal annealing. The pre-growth annealing of the Cu foil and graphene growth were performed in a CVD reactor at 1040˚C. The Cu foil (Alfa Aesar, 99.8%) was annealed in Ar containing $H_2$, and subsequently, the graphene was synthesized on the Cu foil with $CH_4$ diluted to 60-ppm in Ar-$H_2$ gas (see the supplementary information for details). The Cu-foil surface was partially covered with graphene after 20 min. of CVD growth. The post-growth thermal annealing was performed in an ultra-high vacuum (UHV) environment (base pressure $1.0\times10^{-8}$ Pa) for 10 min at temperatures from 200˚C to 700˚C in 100˚C steps. During the annealing, the pressure was kept less than $7.5\times10^{-7}$ Pa.

Transmission electron microscopy (TEM) and electron back scattering diffraction (EBSD) mapping were used to characterize the crystal orientation of the Cu foil. The number of layers, lattice strain, and charge doping in the graphene were evaluated by Raman spectroscopy with a laser excitation wavelength (energy) of 532 nm (2.33 eV). A spot size of the laser was 990 nm, which was estimated from measuring a standard sample with patterned Au on Si substrate.[21] We took Raman mapping data over 20-μm-square areas with 600 nm step-interval. The surface structure of the samples was observed by atomic force microscopy (AFM) and scanning tunneling microscopy (STM).

The as-received Cu foil originally had a polycrystalline structure with micrometer-order grains (see Fig. 1S in the supplementary information), while the Cu grains, after both pre-growth annealing and CVD growth, had two orientations, which were Cu (103) and (001) (see Fig. 2S in the supplementary information). We focused on the CVD-grown graphene on Cu (103) because it is a higher Miller index plane with a unique surface structure. From the AFM height image of the as-grown sample (after CVD growth) shown in Fig. 1(a), we found that only the area covered by CVD-grown graphene had well-aligned facets, and the bare Cu area had no facets. This suggests that the interface interaction between the graphene and Cu plays an important role in inducing this unique faceted structure. Wang *et al.* reported a similar phenomenon in which Cu forms a step structure during cooling just after CVD growth of graphene on it due to the difference in the thermal expansion coefficient between graphene and Cu.[22] We used a combination of a cross-sectional TEM and the selected area electron diffraction (SAED) pattern observations (Fig. 1(b)) in the Cu [010] direction to determine that the facets normal to the Cu foil surface were Cu (103) planes and the inclined facets were (001) planes. This indicated that the observed area was vicinal Cu(103) surface misoriented ~1° toward $[\bar{3}01]$. We also observed that graphene continuously grew across the Cu surface facets. These TEM and SAED results are in good agreement with the EBSD results (Figs. 2S(b) and 2S(c) in the supplementary information).

Next, we investigated the effect of post-growth annealing on the surface morphologies on the Cu(103) facet. Figure 1(c) and (d) show AFM height images of the as-grown and the 500°C post-growth annealed graphene. Both surfaces have faceted structures about 30 nm in height and 500 nm in width. On the Cu (103) facets, triangular patterns of step-terrace structures appeared along $[\bar{3}41]$ and $[36\bar{1}]$ (Fig. 1 (c)) as well as along $[34\bar{1}]$ and $[\bar{3}61]$ (not shown). After the post-growth annealing, the size of the triangular patterns became smaller, indicating that post-growth annealing induced a reorganization of the surface (Fig. 1 (d)).

To investigate the surface structure in detail, we performed STM observations in constant current mode. All of the crystallographic directions in the STM images in Fig. 2(a-d) correspond to those in Fig. 2(e). Low magnification STM (not shown here) revealed that the surface of graphene/Cu(103) had triangular patterns like those observed by AFM (Fig. 1 (c)), after post-growth annealing at 200°C. The higher magnification STM image in Fig. 2(a) shows facets consisting of $[\bar{3}21]$ and $[010]$ directions lined up along the $[34\bar{1}]$ direction. Similar step bands along the $[\bar{3}41]$, $[\bar{3}61]$, $[36\bar{1}]$, and $[010]$ directions were also observed (not shown). This suggests that the triangular patterns that were observed by AFM to be along $[34\bar{1}]$, $[\bar{3}41]$, $[36\bar{1}]$, and $[\bar{3}61]$ in the as-grown graphene correspond to the directions of these step bands. Next, the surface structures after post-growth annealing at different temperatures above 300°C were observed. At 300°C (Fig. 2(b)), the formed facets were tilted by 5.5° from Cu (103) around the [010] axis. They could be assigned to Cu {102} and {104}, as indicated in the atomic side view of Fig. 2(e). At 600°C (Fig. 2(c)), steps along $[\bar{3}41]$, [010] and [30$\bar{1}$] formed. The [$\bar{3}$41] steps were observed as triangular patterns, while [010] and [30$\bar{1}$] steps were as rectangular patterns. The 400˚C and 500˚C post-growth annealing (not shown) gave similar patterns to those at 300˚C and 600˚C, respectively. After the 700˚C post-growth annealing (Fig. 2(d)), the Cu (103) surface showed only rectangular patterns along the [010] and [30$\bar{1}$] directions. As the post-annealing temperatures increased, the original triangular facet patterns changed into different triangular and rectangular patterns, suggesting that the interface energy between the Cu and graphene decreased. The results of STM data in Figs.2(a) and (c) correspond to those of AFM data in Figs.1(c) and (d), respectively. We took each STM data from the same sample but at the different positions because the post-growth annealing and STM measurement were performed in different chambers of the same UHV system. The sample was not exposed to the air during the annealing process and STM measurement. We also confirmed it based on optical micrography and Raman spectroscopy that original CVD-grown graphene remained after the post-growth annealing below 700˚C.

Lattice strain and charge doping in the as-grown and the 500˚C post-growth annealed graphene on Cu (103) were evaluated using Raman spectroscopy. The Raman frequency of the G mode ($\omega_G$) located at about 1580 cm$^{-1}$ depends on charge doping, and that of the 2D mode ($\omega_{2D}$) at about 2700 cm$^{-1}$ is strongly affected by strain.[23, 24] The solid lines in Fig. 3(a) depict the Raman spectra of the G mode and 2D mode averaged from 1089 measurement points which were taken in a 20-μm-square mapping area. More than 80 % of the exposed area was graphene on Cu (103) which was determined from the 500-nm width of Cu (103) and 100-nm width of Cu (001). The dotted lines are fits to each spectrum by using a Voigt curve. The background of the spectra is due to the Cu foil. After the post-growth annealing, the peak became broader, and the height from the base line became lower in both the G mode and the 2D mode, because $\omega_G$ and $\omega_{2D}$ had some variations, as discussed below. Fig. 3(b) plots correlation maps of $\omega_G$ and $\omega_{2D}$ for the as-grown (red circles) and post-annealed (green circles) graphene. The plotted positions of $\omega_G$ and $\omega_{2D}$ were measured at each 1089 measurement point in the Raman mapping. Both data sets can be fitted by a linear correlation $\delta\omega_{2D}/\delta\omega_G \sim 2.2$, which means that the charge-doping level in the graphene was uniform in both samples. The map indicates that the as-grown graphene had a carrier density less than $1\times10^{12}$ cm$^{-2}$ and had compressive strain ($\varepsilon \sim -0.1$ to $-0.3\%$).[22] In contrast, the post-growth annealing uniformly induced hole charge doping in the CVD-grown graphene and increased the carrier density to $4.7\times10^{12}$ cm$^{-2}$. The wider distribution of points for the post-growth annealed graphene indicates inhomogeneous strain over a wide area: from compressive ($\varepsilon \sim -0.3\%$) to tensile ($\varepsilon \sim +0.2\%$). The change in lattice strain due to the post-growth annealing is probably because reorganized small patterns consisting of various Cu crystal planes on the terrace induce different strain levels in the graphene. Each Cu crystal plane has a different atomic arrangement on the surface and gives a different symmetrical relationship and superlattice structure between the graphene and Cu. It has been shown that variations in the Cu crystal planes can induce different doping levels in graphene depending on the work functions of each plane.[9, 14]

In summary, we studied the surface structure of single-layer CVD-grown and post-growth annealed graphene on Cu (103). The graphene/Cu (103) formed a unique faceted structure including Cu (103) and (001), which remained even after the post-growth annealing, while the step-terrace structures with triangular patterns on the (103) facet changed into a combination of smaller triangular and rectangular patterns after the post-growth annealing. The as-grown graphene on Cu (103) had a carrier doping level less than $1\times10^{12}$ cm$^{-2}$, whereas the post-growth thermal annealing induced hole doping at $4.7\times10^{12}$ cm$^{-2}$. In addition, the post-growth annealing affected the strain in graphene. These results indicate that post-growth thermal annealing is a way of manipulating surface structures as well as carrier doping and lattice strain in CVD-grown graphene on Cu.

## Acknowledgments

This work was supported by the JSPS-CNR Bilateral Researcher Exchange Program (Joint Research Projects).

Reference list

1. X. Li, W. Cai, J. An, S. Kim, J. Nah, D. Yang, R. Piner, A. Velamakanni, I. Jung, E. Tutuc, S.K. Banerjee, L. Colombo, and R.S. Ruoff, Science **324**, 1312 (2009).

2. K.S. Kim, Y. Zhao, H. Jang, S.Y. Lee, J.M. Kim, K.S. Kim, J.H. Ahn, P. Kim, J.Y. Choi, and B.H. Hong, Nature **457**, 706 (2009).

3. C. Mattevi, H. Kim, and M. Chhowalla, J. Mater. Chem. **21**, 3324 (2011).

4. J.D. Wood, S.W. Schmucker, A.S. Lyons, E. Pop, and J.W. Lyding, Nano Lett. **11**, 4547 (2011).

5. A.T. Murdock, A. Koos, T.B. Britton, L. Hoube, T. Batten, T. Zhang, A.J. Wilkinson, R.E. Dunin-Borkowski, C.E. Lekka, and N. Grobert, ACS Nano **7**, 1351 (2013).

6. L. Gao, J.R. Guest, and N.P. Guisinger, Nano Lett. **10**, 3512 (2010).

7. Y. Ogawa, B. Hu, C. M. Orofeo, M. Tsuji, K. Ikeda, S. Mizuno, H. Hibino, and H. Ago, J. Phys. Chem. Lett. **3**, 219 (2012).

8. L. Brown, E.B. Lochocki, J. Avila, C.J. Kim, Y. Ogawa, R. W. Havener, D.K. Kim, E.J. Monkman, D.E. Shai, H.I. Wei, M.P. Levendorf, M. Asensio, K.M. Shen, and J. Park, Nano Lett. **14**, 5706 (2014).

9. L. Vitos, A.V. Ruban, H.L. Skriver, and J. Kollár, Surface Science **411**, 186 (1998).

10. P.A. Khomyakov, G. Giovannetti, P.C. Rusu, G. Brocks, J. van den Brink, and P.J. Kelly, Phys. Rev. B **79**, 195425 (2009).

11. A.L. Walter, S. Nie, A. Bostwick, K.S. Kim, L. Moreschini, Y.J. Chang, D. Innocenti, K. Horn,

K.F. McCarty, and E. Rotenberget, Phys. Rev. B **84**, 195443 (2011).

12. O. Frank, J. Vejpravova, V. Holy, L. Kavan, and M. Kalbac, Carbon **68,** 440 (2014).

13. H.I. Rasool, E.B. Song, M. Mecklenburg, B.C. Regan, K.L. Wang, B.H. Weiller, and J.K. Gimzewski, J. Am. Chem. Soc. **133**, 12536 (2011).

14. M. Vondráček, D. Kalita, M. Kučera, L. Fekete, J. Kopeček, J. Lančok, J. Coraux, V. Bouchiat, and J. Honolka, Sci. Rep., **6**, 23663 (2016).

15. Y. Zhang, T. Gao, Y. Gao, S. Xie, Q. Ji, K. Yan, H. Peng, and Z. Liu, ACS Nano, **5**, 4014 (2011).

16. W. Zhu, T. Low, V. Perebeinos, A.A. Bol, Y. Zhu, H.G. Yan, J. Tersoff, and P. Avouris, Nano Lett., 12, 3431 (2012).

17. B. Vasic, A. Zurutuza, and R. Gajic, Carbon 102, 304 (2016).

18. S.S. Chen, Q.Y. Li, Q.M. Zhang, Y. Qu, H X. Ji, R.S. Ruoff, W.W. Cai, Nanotechnology, 23, 365701 (2012).

19. J. Cho, L. Gao, J. Tian, H. Cao, W. Wu, Q. Yu, E.N. Yitamben, B. Fisher, J.R. Guest, Y.P. Chen, and N.P. Guisinger, ACS Nano **5**, 3607 (2011).

20. J.H. Kang, J. Moon, D.J. Kim, Y. Kim, I. Jo, C. Jeon, J. Lee, and B.H. Hong, Nano Lett., 16, 5993 (2016).

21. W. Cai, A.L. Moore, Y. Zhu, X. Li, S. Chen, L. Shi, and R.S. Ruoff, Nano Lett. **10**, 1645 (2010).

22. Z.J. Wang, G. Weinberg, Q. Zhang, T. Lunkenbein, A. Klein-Hoffmann, M. Kurnatowska, M. Plodinec, Q. Li, L. Chi, R. Schloegl, and M.-G. Willinger, ACS Nano, **9**, 1506 (2015).

23. J.E. Lee, G. Ahn, J. Shim, Y. S. Lee, and S. Ryu, Nat. Commun. **3**, 1024 (2012).

24. A. Das, S. Pisana, B. Chakraborty, S. Piscanec, S.K. Saha, U.V. Waghmare, K.S. Novoselov, H.R. Krishnamurthy, A.K. Geim, A.C. Ferrari, and A.K. Sood, Nat. Nanotech. **3**, 210 (2008).

Figure Captions

Fig. 1. (Color) (a) Low-magnification AFM height image of graphene/Cu foil after pre-growth annealing and CVD growth. (b) Cross-sectional TEM image and SAED pattern taken along the facet direction. (c,d). Higher magnification AFM height images taken from (c) as-grown graphene/Cu foil, and (d) after post-growth annealing at 500°C.

Fig. 2. (Color) STM topographic images of graphene/Cu (103) after 10-min of post-growth annealing at (a) 200 °C, (b) 300 °C, (c) 600 °C, and (d) 700 °C. Scan size is 500 nm × 500 nm for all images, and bias voltage and tunneling current are 100 mV and 100 pA. (e) Atomic model and crystalline directions of Cu viewed from [103] (top view) and [010] (side view).

Fig. 3. (Color) (a) Raman spectra averaged over a 20-μm-square Raman maps from as-grown and post-growth annealed (at 500°C) graphene on Cu (103). Solid lines are experimental data, and dashed lines are fits by Voigt curves. Post-growth annealed experimental data and fitted curves are offset to clearly visualize each peak. (b) Correlation map between the Raman frequencies of $\omega_G$ and $\omega_{2D}$ for as-grown and post-annealed (at 500°C) graphene on Cu (103). Dashed lines indicate fitting results.

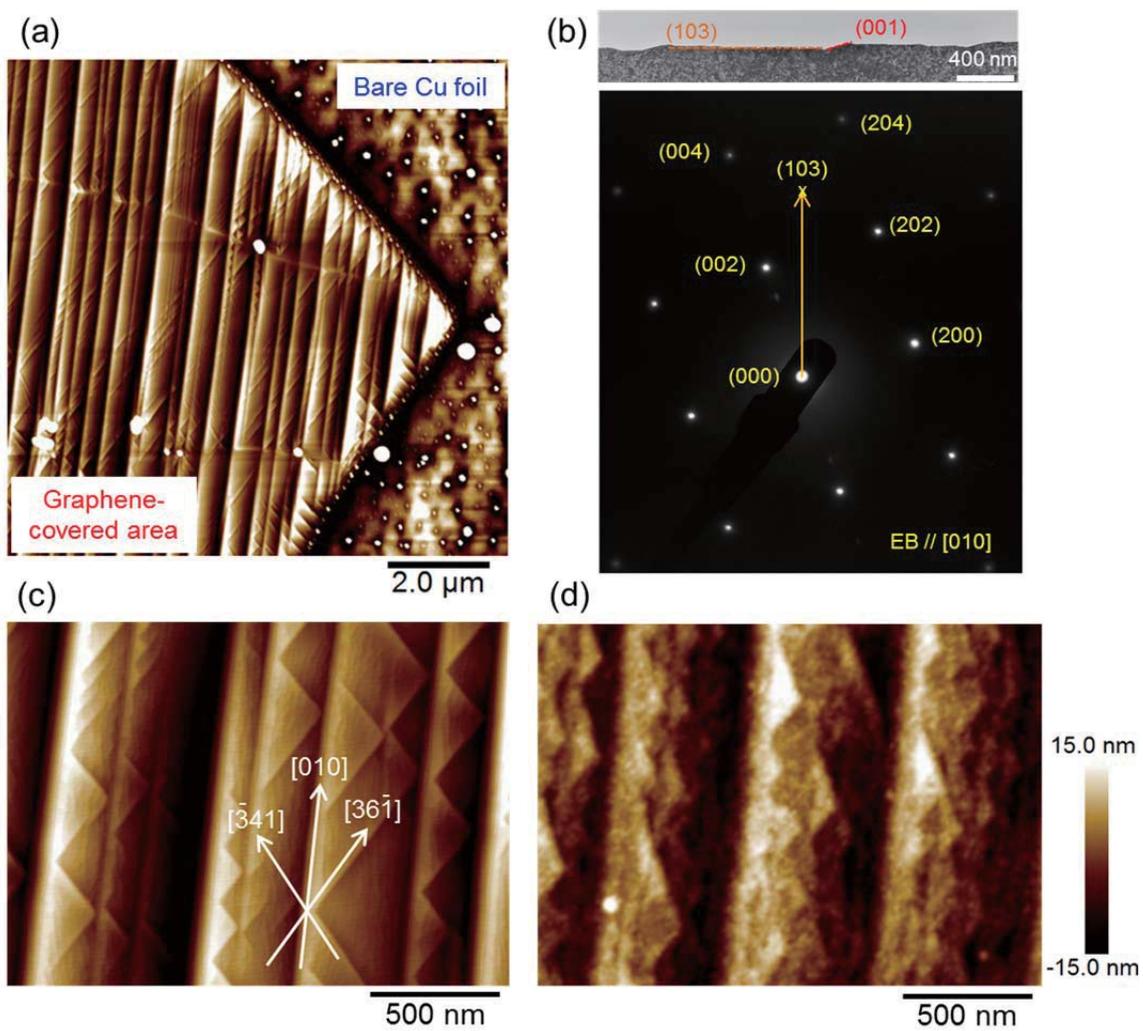

Fig.1. (Color online)

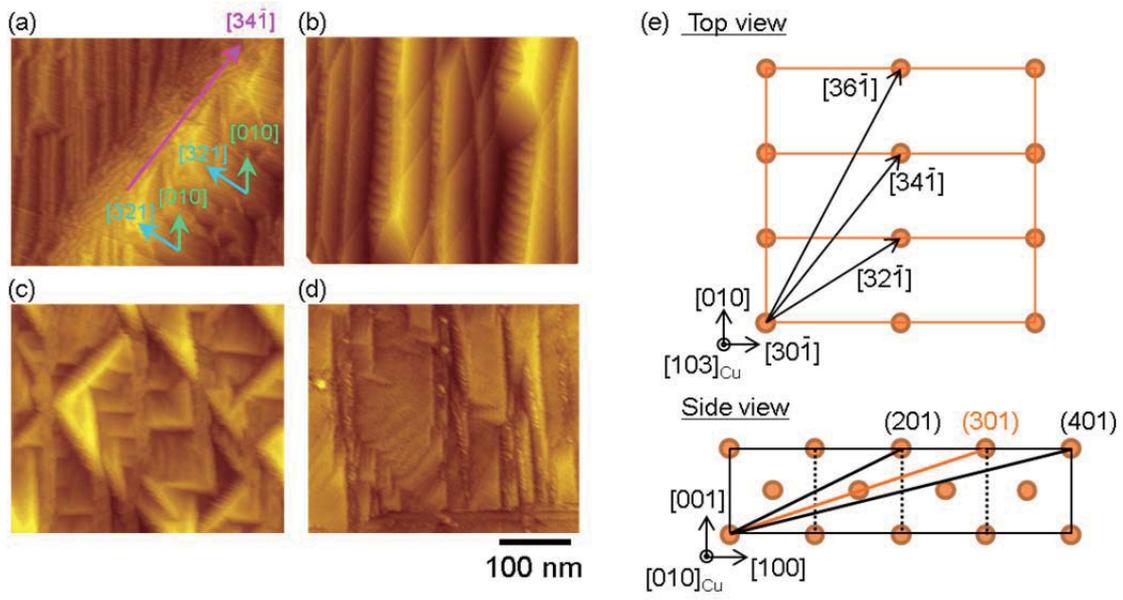

Fig.2. (Color online)

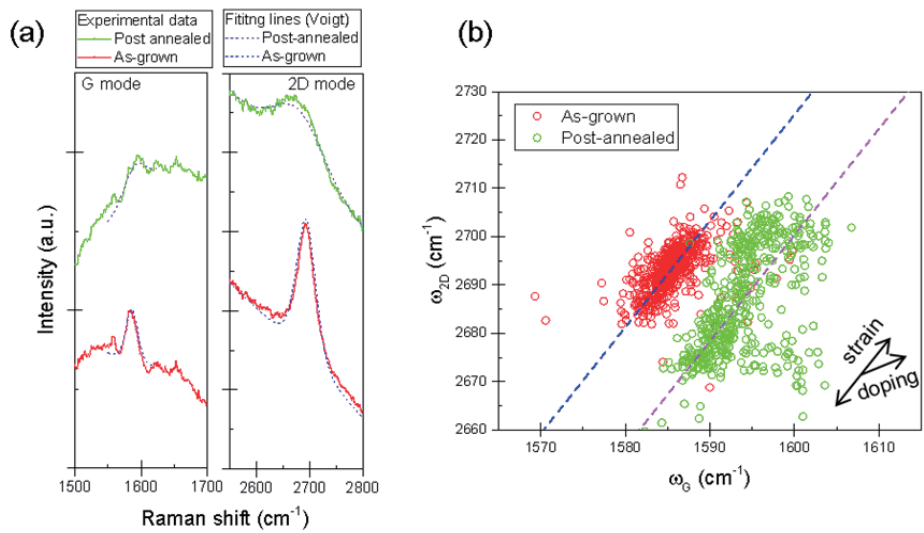

Fig.3. (Color online)

# Surface structures of graphene covered Cu (103)

## *Supplementary information*

**Yui Ogawa[1,*], Yuya Murata[2], Satoru Suzuki[1,†], Hiroki Hibino[1,3], Stefan Heun[2], Yoshitaka Taniyasu[1], and Kazuhide Kumakura[1]**

[1] *NTT Basic Research Laboratories, NTT Corporation,* 3-1, Morinosato Wakamiya, Atsugi, Kanagawa 243-0198, Japan

[2] *NEST, Istituto Nanoscienze-CNR and Scuola Normale Superiore*, Piazza San Silvestro 12, 56127 Pisa, Italy

[3] *Kwansei Gakuin Univ.*, 2-1, Gakuen, Sanda, Hyogo 669-1337, Japan

† Present address: Laboratory of Advanced Science and Technology for Industry, University of Hyogo, Kamigori Hyogo 678-1205, Japan

E-mail: yui.ogawa.er@hco.ntt.co.jp

**Pre-growth annealing and graphene growth:**

As-received commercial Cu foil (Alfa Aesar, 99.8%) was placed on a quartz boat in a quartz tube with a hot-wall-furnace and annealed for 2 hours at 1040°C in an Ar flow of 400 sccm and $H_2$ flow of 100 sccm. Then, graphene was grown at 1040°C in a 1% $CH_4$ (diluted in Ar) flow of 3 sccm, Ar flow of 430 sccm, and $H_2$ flow of 70 sccm. The $CH_4$ concentration was 60 ppm. The pressure was kept at $3.5 \times 10^3$ Pa during the pre-growth annealing and chemical vapor deposition (CVD) growth.

**Crystallographic structure of Cu foils:**

1. As-received Cu foil

The as-received commercial Cu foils had a polycrystalline structure with micrometer-order domains, as shown in the out-of-plane electron back scattering diffraction (EBSD) map in Fig. 1S, left panel. Each color in Fig. S1 indicates a crystal orientation of the Cu grain corresponding to the key orientation map given in the inset at the top right. The right panel shows a unit cell of the face center cubic (FCC) Cu crystal and the orientation of the main low-index Miller planes.

2. Recrystallized Cu foil (after pre-growth annealing and CVD growth)

A scanning electron microscope (SEM) image (Fig. 2S(a)) shows mainly two types of region: homogeneous growth areas at the top left of the image and multiple-boundary areas at the bottom right. The thermal annealing at high temperature, especially during pre-growth annealing, led to a recrystallization of the Cu foil with centimeter-order grains. Figure 2S(b) and (c) show EBSD maps of Cu taken at the top left (homogeneous growth areas) for the out-of-plane and in-plane orientation, respectively. The uniform color in both maps confirms that the Cu was single crystalline without grain boundaries. The inset in Fig. 2S(b) is a pole figure plot, indicating that all measurement points fall at 19.0±0.5 degrees from the (001) plane. Since the angle between the (001) and (103) plane is 18.4 degrees for Cu crystal, the out-of-plane orientation can therefore be assigned to Cu (103). Figure 2S(d) and (e) show EBSD maps taken at the bottom right (multiple-boundary areas) of Fig. 2S(a) for the out-of-plane and in-plane orientation, respectively. These maps indicate that the Cu had a multi-domain structure in which the out-of-plane orientations are aligned to (001) but the in-plane orientations are rotated with respect to each other within 7 degrees.

**Quality of the CVD-grown graphene:**

The quality of the CVD-grown graphene on Cu(103) was evaluated using the D mode of the Raman spectroscopy peak located at 1350 cm$^{-1}$, which originates from defects and dangling bonds in the honeycomb lattice. The D mode was scarcely detected in the entire sample, which indicates low-defect graphene (Fig. S3).

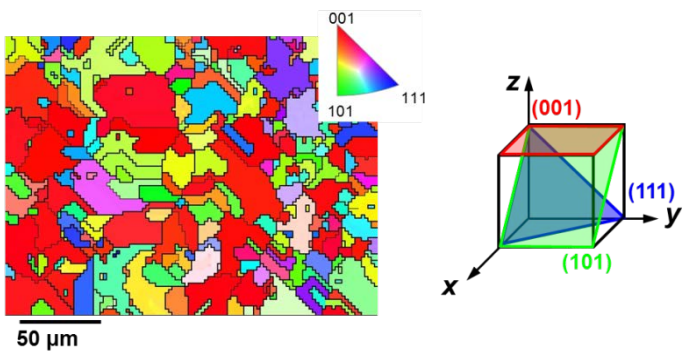

Fig.1S. Left: EBSD map of commercial Cu foil before annealing, and right: key orientation map with unit cell of the Cu FCC lattice.

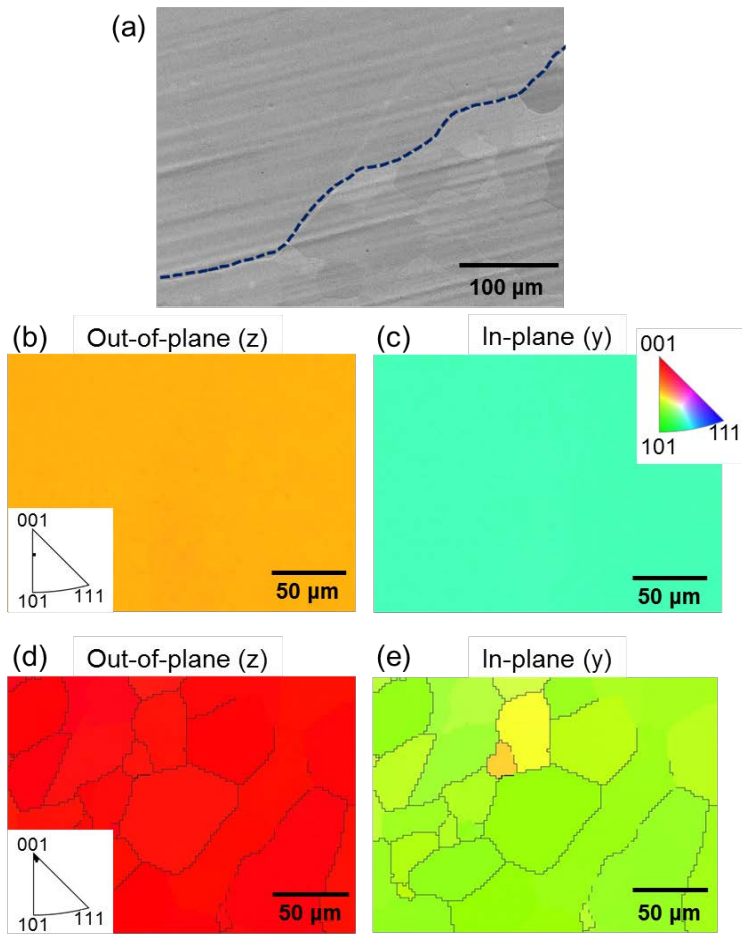

Fig.2S. Recrystallized Cu foil. (a) SEM image of graphene/Cu foil after pre-growth annealing and CVD growth. EBSD maps of two areas in (a) measured at (b,c) top left and (d,e) bottom right. (b,d) and (c,e) indicate out-of-plane (z) and in-plane (y) maps, respectively. Bottom left insets in (b,d) are pole figures of the out-of-plane (z) maps. The inset at the top right in (c) is the key orientation.

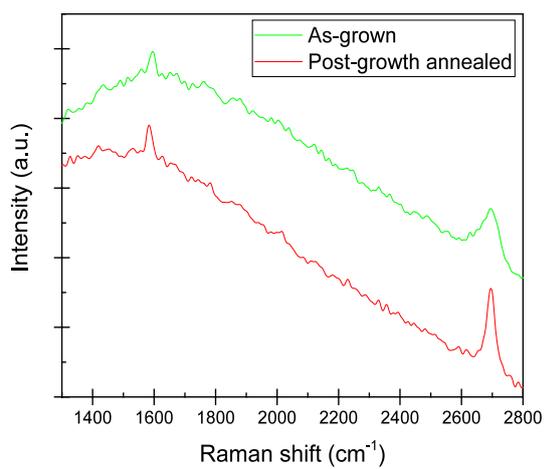

Fig.3S. Typical Raman spectrum measured from as-grown (red) and post-growth annealed (at 500˚C, green) graphene on Cu (103).